\documentclass[twoside,12pt]{article}
\usepackage{obs_study_style}
\usepackage{float}
\usepackage{lmodern}
\usepackage{booktabs}
\usepackage{amsmath,amsfonts,mathtools,bbm,bm}
\usepackage{relsize,exscale}
\usepackage[OT1]{fontenc}
\usepackage[utf8]{inputenc}
\usepackage[linesnumbered,ruled,vlined]{algorithm2e}
\usepackage[inline,shortlabels]{enumitem}
\graphicspath{ {figs/} }

\usepackage{refcount,nameref,zref-xr,zref-user} 
\zxrsetup{toltxlabel} 
\usepackage[hyphens]{url}

\let\href\undefined
\usepackage[colorlinks,citecolor=blue,urlcolor=purple]{hyperref}
\hypersetup{breaklinks=true}

\AtEndDocument{\refstepcounter{theorem}\label{finalthm}}
\AtEndDocument{\refstepcounter{equation}\label{finaleq}}

\newcommand{\E}{\mathbb{E}}

\newcommand{\M}{\mathcal{M}}

\newcommand{\logit}{\text{logit}}

\renewcommand{\Pr}{\mathbb{P}}

\newcommand{\1}{\mathbbm{1}}
\newcommand{\indep}{\mbox{$\perp\!\!\!\perp$}}

\usepackage[dvipsnames]{xcolor}

\newcommand{\authorlist}{%
  \name Nima S.~Hejazi \email nhejazi@hsph.harvard.edu\\
  \addr Department of Biostatistics,\\
  T.H.~Chan School of Public Health,\\
  Harvard University\\
  655 Huntington Ave.,\\
  Boston, MA 02115
  \AND
  \name Mark J.~{van der Laan} \email laan@berkeley.edu\\
  \addr Division of Biostatistics,\\
  School of Public Health,\\
  University of California, Berkeley\\
  2121 Berkeley Way,\\
  Berkeley, CA 94720
}

\newcommand{\titlepaper}{Revisiting the Propensity Score's Central Role:
Towards Bridging Balance and Efficiency in the Era of Causal Machine Learning
}

\heading{X}{2022}{}{}{}{Nima S.~Hejazi and Mark J.~van der Laan}

\ShortHeadings{Revisiting The Role of the Propensity Score in Causal
  Machine Learning}{Hejazi and van der Laan}
\firstpageno{1}

\begin{document}
\title{\titlepaper}
\author{\authorlist}
\maketitle

\begin{abstract}
About forty years ago, in a now--seminal
contribution,~\citet{rosenbaum1983central} introduced a critical
characterization of the propensity score as a central quantity for drawing
causal inferences in observational study settings. In the decades since, much
progress has been made across several research frontiers in causal inference,
notably including the re-weighting and matching paradigms. Focusing on the
former and specifically on its intersection with machine learning and
semiparametric efficiency theory, we re-examine the role of the propensity score
in modern methodological developments. As~\citet{rosenbaum1983central}'s
contribution spurred a focus on the balancing property of the propensity score,
we re-examine the degree to which and how this property plays a role in the
development of asymptotically efficient estimators of causal effects; moreover,
we discuss a connection between the balancing property and efficient estimation
in the form of score equations and propose a score test for evaluating whether
an estimator achieves empirical balance.
\end{abstract}


\section{The Propensity Score's Central Role: A Retrospective}

In observational studies that aim to evaluate the causal effects of well-defined
interventions~\citep[e.g.,][]{hernan2008does,pearl2010brief}, a key inferential
obstacle arises in the form of potential confounding of the treatment--response
relationship by baseline (pre-treatment) covariates. Unlike observational
studies, randomized controlled trials (RCTs) feature a built-in safeguard
against this form of confounding --- specifically, since the treatment's
allocation among study units is, on average, balanced across strata defined by
the baseline covariates, confounding of the treatment--response relationship is
theoretically expected to be a non-issue. It is this inferential safeguard that
is in part responsible for RCTs being considered as ``gold standard'' tools for
generating evidence in biomedical and health research.
As observational studies lack any such built-in protective measure, significant
care is required, in both the design and analysis stages, to mitigate
confounding by addressing systematic differences between treatment and control
groups. Such considerations come in the form of assumptions like that of strong
ignorability (an observational analog to the randomization assumption) and
adjustment for baseline covariates when estimating nuisance parameters critical
for the construction of estimators of the causal effects of
interest.~\citet{rosenbaum1983central} introduced the propensity score (the
probability of receiving treatment conditional on baseline covariates) and
related this quantity to the issue of balance between treatment groups. In their
contribution, these authors outline the propensity score's membership in a class
of \textit{balancing scores}, which satisfy conditional independence between
treatment assignment and baseline covariates. Through their analysis, these
authors relate this form of conditional independence to the (untestable) strong
ignorability assumption and show that the propensity score plays a critical role
in causal inference, with uses in matching study units, in stratification for
weighting-based adjustment, and in covariance adjustment in analyses centered on
the general linear model. On account of the propensity score's close
correspondence with the notion of balance, a sharp focus has been placed on this
particular property in the decades since~\citet{rosenbaum1983central}'s
contribution, yet most ideas have been restricted to viewing balance in terms of
the properties of parametric modeling-based estimators of the propensity score.
In this case, the balancing property of a propensity score estimator is the
finite-sample analogue of the conditional independence assumption that
theoretically defines a balancing score. Taking a view rooted outside the
traditional culture of parametric modeling, we define a notion of balance for an
estimator of the propensity score, with respect to particular functions of
baseline covariates, showing that this function-specific balancing property
corresponds with solving function-specific score equations, connecting this
balancing property with formal criteria for asymptotically efficient estimation.

To formalize our arguments, consider an observational study that collects data
on $n$ units, $O_1, \ldots, O_n$, sampled independently and identically from an
(unknown) distribution $P_0$, that is, $O \sim P_0 \in \M$, where $\M$ is
a realistic (nonparametric) statistical model placing only minimal and plausible
restrictions on the form of $P_0$. The data available on the
\textit{i}\textsuperscript{th} unit $O_i$ may be partitioned by time-ordering as
$(X_i, Z_i, R_i)$, for a vector of baseline covariates $X$, a binary treatment
$Z \in \{0, 1\}$, and a response $R$ (where possible, we borrow the notation
of~\citet{rosenbaum1983central} in homage). For any unit $O$, $R(0)$ and $R(1)$
are the \textit{potential outcomes}~\citep{neyman1938contribution,
rubin1978bayesian, rubin2005causal} of $R$, mutually unobservable quantities
that arise when the treatment $Z$ takes the values $Z = 0$ and $Z = 1$,
respectively. The population average treatment effect (ATE) is $\tau_0
= \E_0[R(1) - R(0)]$, a difference of counterfactual means; the naught subscript
refers to the true distribution $P_0$. Under standard assumptions of consistency
($R(1) \equiv R \mid Z = 1$)~\citep{pearl2010brief}, lack of
interference~\citep{cox1958planning}, positivity of treatment assignment
($\delta < \Pr(Z = 1 \mid X) < 1 - \delta$ for $\delta > 0$), and strong
ignorability of treatment assignment (i.e., randomization of treatment
assignment in RCTs), $\tau_0$ is identifiable; furthermore, it is estimable by
either substitution (plug-in) or inverse probability weighted (IPW) estimators,
which arise from differing identification strategies. A substitution estimator
takes the form $\tau_n^{\text{SUB}} = \E_{n,X} \{\overline{Q}_n(1, X)
- \overline{Q}_n(0, X)\}$, where $\overline{Q}_n(Z, X) \coloneqq \E_n(R \mid Z,
X)$ is an estimator of the true response mechanism's conditional mean
$\overline{Q}_0(Z, X) \coloneqq \E_0(R \mid Z, X)$, while an IPW estimator is
$\tau_n^{\text{IPW}} = \E_{n,X} \{[\1(Z = 1) / e_n(X) - \1(Z = 0) / (1
- e_n(X))] R\}$, where $e_n(X) \coloneqq \Pr_n(Z = 1 \mid X)$ is an estimator of
the true propensity score $e_0(X) \coloneqq \Pr_0(Z = 1 \mid X)$. In both cases,
expectations w.r.t.~$X$ are simply computed via an empirical mean. Throughout,
as with the naught subscript, the subscript $n$ denotes estimates based on the
empirical distribution $P_n$ of the sampled units $O_1, \ldots, O_n$. On
occasion, we will rely upon standard notation from empirical process theory;
specifically, we will let $P f \coloneqq \int f(O) dP$ and $P_n f \coloneqq
n^{-1} \sum_{i=1}^n f(O_i)$ whenever convenient.

Salient to arguments about the balancing property, IPW estimators may be viewed
as achieving balance across treatment conditions by upweighting (or
downweighting) units from strata of $X$ that are underrepresented (or
overrepresented) in each of the treatment groups, resulting in an artificially
constructed pseudo-population~\citep{horvitz1952generalization,
hernan2021causal} in which treatment assignment $Z$ is empirically marginally
independent of (or ``balanced on'') $X$. That is, inverse probability weighting
by treatment propensity, using $e_0(X)$, creates a hypothetical
pseudo-population in which empirically $X \indep Z$ (essentially by creating
copies of underrepresented units), allowing for inference on the effect of $Z$
on $R$ without confounding by $X$. This form of balance corresponds with the
(conditional) independence condition of~\citet{rosenbaum1983central} ($X \indep
Z \mid e_0(X)$), which clarifies that the propensity score $e_0(X)$ can be used
to mitigate confounding of the $Z$--$R$ relationship by $X$, by enforcing
conditional independence of $X$ and $Z$ given $e_0(X)$.
\citet{rosenbaum1983central} use this conditional independence relationship to
theoretically characterize balancing scores, defining a balancing score $b_0(X)$
as satisfying the condition $X \indep Z \mid b_0(X)$. The propensity score is
itself a balancing score, $e_0(X) = f(b_0(X))$ --- in fact, it is the
``coarsest'' among this class of scores in the sense of minimally satisfying
this form of independence. We emphasize that the empirical analogue of this
balancing property applied to an estimator $e_n(X)$ of $e_0(X)$ drives the
efficiency of estimators of the causal effect of interest; for example, IPW
estimators using $e_0(X)$ are consistent and asymptotically linear but very
inefficient \citep[e.g.,][]{vdl2003unified}.
While the theoretical characterization of balance intuitively explains how this
property is central to constructing consistent estimators of causal effects, it
ignores a property generally desirable in any estimator --- efficiency.


\section{The Empirical Balance--Efficiency Tradeoff}\label{balance_effic}

Although the propensity score's use in inducing empirical balance between
treatment groups has intuitive appeal, classical estimators relying solely on
this property for a specific parametric model, as traditionally formulated,
generally fail to achieve asymptotic efficiency, even when the parametric model
for the propensity score is correctly specified. In fact, it has been
established that neither the IPW estimator nor the propensity score
stratification-based estimator (proposed by~\citet{rosenbaum1983central}) are
generally asymptotically efficient. Drawing upon semiparametric efficiency
theory, we note that the \textit{efficient influence function} (EIF) arises as
a central object, uniquely represented as the canonical gradient of the pathwise
derivative of the target parameter ($\tau_0$) at a distribution $P$ in the model
$\M$~\citep[see, e.g.,][]{bickel1993efficient, vdvaart2000asymptotic,
vdl2003unified, tsiatis2007semiparametric, kennedy2016semiparametric,
hines2022demystifying}. The EIF's importance comes from the fact that it
characterizes the best possible asymptotic variance, or nonparametric efficiency
bound, among all regular asymptotically linear estimators of a target parameter.
For this reason, the EIF is commonly used as a critical ingredient in strategies
for the construction of efficient estimators. For example, efficient estimation
frameworks popular in modern practice --- such as one-step
estimation~\citep{pfanzagl1985contributions, bickel1993efficient}, estimating
equations~\citep{vdl2003unified, bang2005doubly}, and targeted minimum loss
estimation~\citep{vdl2006targeted, vdl2011targeted} --- construct estimators by
unique updating procedures that each reference the form of the EIF in different
ways, resulting in candidate estimators with desirable asymptotic behavior.
Under standard regularity conditions, an estimator $\tau_n$ of the target
parameter $\tau_0$ is asymptotically linear when
\begin{equation}\label{eqn:ral}
  \tau_n - \tau_0 = \frac{1}{n} \sum_{i = 1}^n D^{\star}(P_0)(O_i) +
    o_p(n^{-1/2}),
\end{equation}
where $D^{\star}(P_0)$ is the EIF at the true data-generating distribution
$P_0$. An asymptotically linear estimator $\tau_n$ is generally (asymptotically)
efficient when it solves the EIF estimating equation, i.e., $P_n D^{\star}(P_0)
\approx 0$, in which case $\tau_n$ has limit distribution $\text{N}(\tau_0,
P \{D^{\star}(P_0)\}^2)$ with asymptotic variance matching that of the EIF
$D^{\star}(P_0)$. As such, representations of the EIF play a key role in
constructing efficient estimators.

\subsection{Score Equations Characterize Asymptotic
Efficiency}\label{score_effic}

In causal inference problems, the EIF is indexed by at least two nuisance
quantities, among which the propensity score invariably appears. It is in this
way that $e_0(X)$ plays a key role in the construction of efficient estimators.
Going forward, to simplify notational burden, we focus on a single component of
the ATE, the counterfactual mean of the response under treatment $Z = 1$, i.e.,
$\tau_0 \coloneqq \E_0 R(1)$. The EIF $D^{\star}(P)$ of $\tau_0$ at the
data-generating distribution $P_0 \in \M$ may be expressed
\begin{equation}\label{eqn:eif_sub}
  D^{\star}(P_0)(O) = \frac{\1(Z = 1)}{e_0(X)} (R -
    \overline{Q}_0(Z, X)) + \overline{Q}_0(1, X) - \tau_0 \ .
\end{equation}
The form of expression~\eqref{eqn:eif_sub} for the EIF of $\tau_0$ is rather
instructive, revealing that the propensity score enters into a score term for
the response mechanism $h(X) (R - \overline{Q}_0(Z, X))$, with $h(X) = \1(Z
= 1) / e_0(X)$, as an inverse weight applied to a residual for the conditional
mean of the response given $(Z, X)$. This expression for the EIF places emphasis
on estimation of the response mechanism $\overline{Q}_0(Z, X)$, highlighting
that an efficient estimator must incorporate a nuisance estimator
$\overline{Q}_n(Z, X)$ that suitably solves the EIF estimating equation, $P_n
D^{\star}(\overline{Q}_n, e_n) \approx 0$. Owing to this emphasis on
$\overline{Q}_0(Z, X)$, such expressions for the EIF are most amenable to the
construction of efficient estimators relying upon the substitution formula, such
as in targeted minimum loss estimation~\citep{vdl2011targeted}, which updates
initial estimates of $\overline{Q}_n$ via a one-dimensional parametric update
step that depends on the EIF's form.

Unlike their substitution-based counterparts, IPW estimators rely
\textit{solely} on the propensity score. Saliently, an alternative expression
for the EIF --- the Augmented IPW (AIPW) representation
of~\citet{robins1992recovery, robins1995semiparametric} --- is more suitable for
their characterization:
\begin{equation}\label{eqn:eif_ipw}
  D^{\star}(P_0)(O) = \underbrace{\frac{\1(Z = 1)}{e_0(X)} R -
    \tau_0}_{D_{\text{IPW}}} - \underbrace{\frac{\overline{Q}_0(1, X)}{e_0(X)}
    (Z - e_0(X))}_{D_{\text{CAR}}} \ .
\end{equation}
The AIPW representation~\eqref{eqn:eif_ipw} of the EIF can be shown to be
equivalent to expression~\eqref{eqn:eif_sub} but stresses instead the importance
of the propensity score through the score term for the treatment mechanism
$h(X) (Z - e_0(X))$, with $h(X) = \overline{Q}_0(1, X) / e_0(X)$. As indicated,
the form of~\eqref{eqn:eif_ipw} admits a decomposition: the first term
$D_{\text{IPW}}$ is the IPW estimating equation (the mapping defining these
Z-estimators) while the second term $D_{\text{CAR}}$ is a projection (of
$D_{\text{IPW}}$ onto the space of all functions of $(Z,X)$ that are mean-zero
conditional on $X$) satisfying coarsening-at-random
(CAR)~\citep{vdl2003unified}.
Since IPW estimators are defined as solutions to $P_n D_{\text{IPW}}(e_n)
\approx 0$, this term is trivially solved by construction of $\tau_n$; then,
expression~\eqref{eqn:eif_ipw} states that an efficient IPW-type estimator must
be a solution to $P_n D_{\text{CAR}}(e_n, \overline{Q}_n) \approx 0$ principally
through a suitable estimator of the propensity score $e_n(X)$, suggesting that
one should prioritize an estimator $e_n(X)$ that satisfies this criterion when
seeking an efficient estimator $\tau_n$ of $\tau_0$. Again, note that, even when
$e_0(X)$ is known, using $e_0(X)$ would fail to solve relevant score equations,
including $P_n D_{\text{CAR}}(e_n,\overline{Q}_n)\approx 0$. Together,
expressions~\eqref{eqn:eif_sub} and~\eqref{eqn:eif_ipw} provide criteria for the
construction of efficient estimators --- notably, both depend on proper
estimation of the propensity score $e_n(X)$ in the  sense that it solves key
score equations.

\subsection{Score Equations Characterize Directional Empirical
Balance}\label{score_balance}

From expressions~\eqref{eqn:eif_sub} and~\eqref{eqn:eif_ipw}, we have seen that
score terms of the form $h(X)(R - \overline{Q}_0(Z,X))$ and $h(X)(Z - e_0(X))$,
for particular choices of the weighting function $h(X)$, play critical roles in
criteria for asymptotic efficiency. We now argue that such score terms also
characterize the empirical balancing property, though this has not been, to the
best of our knowledge, extensively explored to date. Recall
that~\citet{rosenbaum1983central} characterize the theoretical balancing
property in terms of a statistical conditional independence condition $X \indep
Z \mid e_0(X)$. This view on the balancing property has motivated the
development of a host of diagnostic procedures to evaluate the sample-level
balance provided by candidate estimators $e_n(X)$, with graphical procedures
(e.g., the ``Love plot'') enjoying much popularity and software implementations
often accruing many thousands of downloads (e.g.,~\citet{greifer2022cobalt}'s
\texttt{cobalt} \texttt{R} package). Yet, the popularity of such approaches
belies their scientific and statistical value: these diagnostic techniques are
limited to revealing only whether an estimator $e_n(X)$ induces empirical
balance with respect to the statistical model underlying the estimator. For
example, logistic regression remains an exceedingly popular candidate estimator
of the propensity score but intrinsically assumes the (logit of) the conditional
mean of $Z$ given $X$ to be adequately described as a linear function of $X$. Of
course, this amounts to a significant (and often unrealistic) restriction on the
statistical model $\M$. Much worse though, this assumption is usually imposed
only for the sake of mathematical convenience, hardly ever motivated by domain
knowledge. Even when the parametric model for the propensity score is correct,
these techniques check only the degree to which a maximum likelihood estimator
$e_n(X)$ satisfies empirical balance within restrictive (small) statistical
models, thereby only achieving empirical balance $Z \indep f(X)$, given
$e_n(X)$, for a very limited set of functions $f(X)$.
In so doing, the balancing property w.r.t.~the chosen parametric model is
emphasized over such fundamental concerns as the efficiency of the estimator
$\tau_n$, ignoring even the fact that the resulting estimator will generally
fail to even achieve consistency.

Fortunately, focusing on univariate or multivariate balance under parametric
modeling assumptions --- convenient as it may be --- is hardly the only option.
When the propensity score estimator $e_n(X)$ is selected as a solution to score
equations of the form $P_n s(e_n; f) \approx 0$, for $s(Z, X; f) = f(X)(Z
- e_0(X))$, an $f$-specific form of conditional independence is satisfied,
$Z \indep f(X) \mid e_n(X)$. Lemma~\ref{lem:balance} summarizes this.
\begin{lemma}[Score-based Balance]\label{lem:balance}
Let $\mathcal{F}$ contain a rich class of functions and, for an arbitrary $f
\in \mathcal{F}$, define scores of the form $s(Z, X; f) = f(X)(Z - e_0(X))$.
When a corresponding score equation $P_n s(e_n; f) \approx 0$ is solved for
a given $f$, the null hypothesis $H_0(f): \E_0(Z \mid f(X), e_n(X)) = \E_0(Z
\mid e_n(X))$ holds, as the data provide no signal against $H_0(f)$; moroever,
no valid test of $H_0(f)$ will reject this null hypothesis. When $H_0(f)$ holds,
$f(X)$ contains no information, beyond that captured by $e_n(X)$, useful for
predicting treatment status $Z$ from covariates $X$ --- that is, the empirical
balance induced by $e_n(X)$ cannot be improved by $f(X)$.
\end{lemma}
Lemma~\ref{lem:balance} provides a score-based criterion for characterizing the
empirical balancing property and frames its evaluation in terms of a class of
hypothesis tests.
When a sequence of such tests uniformly fails to reject a family of $f$-specific
null hypotheses $\{H_0(f) : f \in \mathcal{F}\}$, there is no empirical evidence
to contradict the equality under the null for the family of $f \in \mathcal{F}$;
then, no such $f(X)$ contains information about $Z$ not already captured by
$e_n(X)$, implying that $e_n(X)$ enforces balance in this $f$-specific sense.
Since $P_n s(e_n; f)$ may be viewed as a measure of the degree to which $Z$ is
independent of $f(X)$, given $e_n(X)$, this $f$-specific empirical balance may
be enforced via score tests for $H_0(f)$ by selecting $e_n(X)$ so as to ensure
$P_n s(e_n; f) \approx 0$.

To employ such a hypothesis testing strategy, consider a model $\logit[\E_0(Z
\mid X)] = \logit[e_n(X)] + \beta f(X)$ for $f \in \mathcal{F}$, in which
$e_n(X)$ is taken as an offset. Under this assumption, the null hypothesis may
be reframed, for fixed $f \in \mathcal{F}$, as $H_0(f): \beta = 0$ against the
alternative $H_1(f): \beta \neq 0$, so that a hypothesis test of this form
corresponds to testing the null of independence. Basing such a hypothesis test
on the score of the empirical log-likelihood, given by $\E_n f(X)(Z - e_n(X))$,
at $\beta = 0$ leads to a score test that simply evaluates the magnitude of this
score as a test statistic, rejecting the null hypothesis when it moves
appreciably far from zero. When the estimator $e_n(X)$ solves the score equation
$\E_n f(X)(Z - e_n(X)) \approx 0$, there cannot be evidence against $H_0(f)$. As
noted above, when $\mathcal{F}$ is a rich class, and this type of score is
satisfied for many $f(X)$, then this score test fails to reject the null
hypothesis $H_0(f): \beta = 0$ in the class $\mathcal{F}$, implying that
$\E_0(Z \mid f(X), e_n(X)) = \E_0(Z \mid e_n(X))$ --- that is, $X \indep Z \mid
e_n(X)$. Finally, when $\mathcal{F}$ is rich enough to include weighting
functions appearing in the EIF, e.g., $\overline{Q}_n(1,X) / e_n(X)$ in
expression~\eqref{eqn:eif_ipw}, then $e_n(X)$ will both enforce balance in $X$
over $\mathcal{F}$ and the resultant estimator $\tau_n$ will be asymptotically
efficient on account of satisfying the EIF estimating equation $P_n
D_{\text{CAR}}(e_n, \overline{Q}_n) \approx 0$. Of course, satisfying this
score-based balancing criterion for many $f \in \mathcal{F}$, providing balance
over $X$ w.r.t.~the linear span of these $f$, may also automatically solve the
score equation for the $f(X)$ appearing in the relevant EIF estimating equation.

\section{Statistical Techniques for Solving Score Equations}\label{score_solve}

To this point we have reviewed the critical role that score equations play in
asymptotic efficiency, through their appearance in the EIF estimating equation,
and outlined their connection to the notion of empirical balance proposed
by~\citet{rosenbaum1983central} via Lemma~\ref{lem:balance} and our proposed
score test. Owing to their critical role in constructing efficient estimators,
several classes of techniques for solving score equations have been proposed at
the interface of causal machine learning and semiparametric efficiency theory.
We next selectively review two successful frameworks that have been applied to
this end: targeted minimum loss estimation~\citep{vdl2006targeted} and
nonparametric sieve estimation.

\subsection{Targeted Updating of the Propensity Score Estimator}

The targeted minimum loss estimation (or targeted learning) framework focuses on
constructing efficient substitution estimators, and, as such, features
a targeting (or updating) step usually applied only to initial estimates of the
response mechanism $\overline{Q}_n(Z,X)$. For example, a TML estimator
$\tau^{\star}_n$ of the counterfactual mean $\tau_0$ is constructed in two
steps, by, first, generating an initial estimate $\overline{Q}_n(Z,X)$ (for
which flexible machine learning strategies~\citep[e.g.,][]{vdl2007super} are
recommended) and, second, perturbing the initial estimate using a univariate
parametric model $\logit(\overline{Q}_n^{\star}(Z,X))
= \logit(\overline{Q}_n(Z,X)) + \epsilon h(X)$, where $h(X) = \1(Z = 1)
/ e_n(X)$ as in expression~\eqref{eqn:eif_sub}. This correction ensures that
$\overline{Q}_n^{\star}(Z,X)$ is free of ``plug-in bias,'' allowing the TML
estimator $\tau^{\star}_n$ to achieve asymptotic linearity by acting as an
approximate solution to the EIF estimating equation, making the TML estimator
$\tau^{\star}_n$ asymptotically efficient. In principle, this same perturbation
strategy may be applied to updating propensity score estimators $e_n(X)$ so as
to enforce balance (or other such desiderata) while still ensuring that the
estimators remain as solutions to the EIF estimating equation. We note that the
targeted learning framework has been, and continues to be, the subject of
fervent research~\citep[][]{vdl2011targeted,vdl2018targeted}, and that there are
many variations of TML estimators suited to different goals, even including
those adjusting specifically for balancing scores~\citep{lendle2015balancing}.

In considering alternative applications of the targeted updating approach,
\citet[][see Theorem 1]{vdl2014targeted} proposed a procedure for the
construction of targeted IPW estimators. Such IPW estimators are based on
applying the one-dimensional parametric update $\logit(e_n^{\star}(X))
= \logit(e_n(X)) + \epsilon h(X)$ to map an initial propensity score estimator
$e_n(X)$ into an updated version $e_n^{\star}(X)$ so as to satisfy asymptotic
linearity, though asymptotic efficiency remained elusive.
As IPW estimation avoids explicit modeling of the response mechanism,
constructing efficient estimators is both philosophically and technically
challenging, as such estimators must solve the $D_{\text{CAR}}$ component of the
EIF, which includes the response mechanism in the numerator of the inverse
weight of the score term, i.e., $h(X) = \overline{Q}_n(1,X) / e_n(X)$ (as in
expression~\eqref{eqn:eif_ipw}). One may, for example, take an approach based
in universal least favorable update models~\citep{vdl2016one}, tracking updates
to the estimator $e_n^{\star}(X)$ locally along a path in a one-dimensional
parametric model formulated to solve the score equation $P_n h(X)(Z - e_n(X))
\approx 0$.
To circumvent modeling the response mechanism,~\citet{vdl2014targeted} proposed
instead replacing $\overline{Q}_n(1, X)$ with an estimator of an artificial
nuisance parameter $\overline{Q}_0^r(1, X) \coloneqq \E_0(R \mid Z,
\bar{e}_n(X))$, in which $\bar{e}_n(X)$ is a fixed summary measure of the
covariates $X$. Constructing such artificial nuisance quantities requires
mathematically sophisticated dimension reduction efforts, that, while difficult
to generalize, avoid the logical contradiction associated with directly modeling
the response $R$. Theoretical investigations revealed these targeted IPW
estimators to satisfy asymptotic linearity but fail to satisfy standard
regularity conditions, sharply limiting their practical use. These targeted IPW
estimators were implemented and are available in the \texttt{drtmle} \texttt{R}
package~\citep{benkeser2022doubly}.

\subsection{Undersmoothing of the Propensity Score Estimator}

As we have seen, asymptotically efficient estimators can only be constructed as
solutions to score equations (e.g., $P_n D_{\text{CAR}}(e_n, \overline{Q}_n)
\approx 0$) rooted in careful study of the EIF. In the context of IPW
estimation, these scores take the general form $s(Z, X; h) = h(X)(Z - e_n(X))$,
where $h(X)$ is an appropriate weighting function, e.g., $h(X)
= \overline{Q}_n(1,X) / e_n(X)$ when $\tau_0$ is the counterfactual mean under
treatment. When $e_n(X)$ is constructed by way of a sieve MLE (or NP-MLE, when
such exists), undersmoothing may be applied to this initial estimator so as to
ensure it solves relevant score equations, i.e., $P_n s(e_n; h) \approx 0$. For
such efforts to succeed, not only must $e_n(X)$ converge to $e_0(X)$ at
a suitably fast rate (i.e., rate-consistency), but $e_n(X)$ must act as an MLE
over the set of statistical models over which the sieve is applied. In recent
work, \citet{ertefaie2022nonparametric} demonstrated, both theoretically and
practically, that undersmoothing of a highly adaptive lasso (HAL) estimator of
$e_n(X)$ allows for the construction of asymptotically linear and efficient IPW
estimators of $\tau_0$. Relatedly, \citet{vdl2019efficient} proved that
substitution estimators using undersmoothed HAL are efficient as well, which,
together with~\citet{ertefaie2022nonparametric}, demonstrates the utility of HAL
across two important and popular classes of estimators. The properties of these
two HAL-based estimators of causal effects rely on key properties of the HAL
estimator itself, first investigated
by~\citet{vdl2015generally,vdl2017generally}, who showed HAL to be an MLE for
a model $\M_{\lambda_0}$ indexed by the true sectional variation norm
$\lambda_0$ of the HAL representation of a target functional (e.g., $e_{0,
\lambda_0}(X)$). The HAL estimator assumes that the target functional, $e_{0,
\lambda_0}(X)$ in this context, to be c\`{a}dl\`{a}g (i.e., RCLL) and of bounded
sectional variation norm ($\lambda_0 < \infty$); this estimator is available in
the open source \texttt{hal9001} \texttt{R}
package~\citep{coyle2021hal9001-rpkg, hejazi2020hal9001-joss}. Using a HAL-MLE
for $e_{n, \lambda_n}(X)$, \citet{ertefaie2022nonparametric} outlined conditions
and formulated selection procedures for appropriately undersmoothing $e_{n,
\lambda_n}(X)$ such that the resultant IPW estimator
$\tau_{n,\lambda_n}^{\text{IPW}}$ would attain the nonparametric efficiency
bound, solving the EIF estimating equation by acting as a solution to $P_n
D_{\text{CAR}}(e_{n,\lambda_n}, \overline{Q}_n) \approx 0$. Contemporaneously,
\citet{hejazi2022efficient} also investigated novel undersmoothing selection
criteria agnostic to the EIF's form for HAL-based estimators of the generalized
propensity score~\citep{imbens2000role, hirano2004propensity}.

Prior work has considered the undersmoothing of propensity score
estimators, notably including~\citet{hirano2003efficient}, who proposed the use
of a logistic series estimator for $e_n(X)$. While these authors showed
undersmoothing of their $e_n(X)$ estimator to be capable of generating an
efficient IPW estimator of the ATE, their approach is practically limited by
requiring the propensity score $e_0(X)$ to be a (highly smooth) $k$-times
differentiable function of $X$. What's more, this and related approaches fail to
appropriately emphasize that efficiency of the estimator $\tau_n$ is a direct
result of $e_n(X)$ solving the score equation arising from the EIF. By contrast,
the more recent developments of~\citet{ertefaie2022nonparametric} highlight this
point; furthermore, based on our Lemma~\ref{lem:balance}, a natural extension of
the latter approach would be to focus on solving a broad range of score
equations (of the form $P_n h(X)(Z - e_{n, \lambda_n}(X)) \approx 0$),
ultimately resulting in both the downstream estimator $\tau_n$ attaining
asymptotic efficiency and the propensity score estimator $e_n(X)$ inducing
empirical balance without reliance on brittle modeling assumptions. Of course,
our proposed score test implies that solving a range of score equations yields
an estimator satisfying the empirical balancing property, positioning
undersmoothing as a potentially critical tool for achieving empirical balance
and, thereby, asymptotic efficiency. Nevertheless, in finite samples, beyond
using HAL, one may wish to enforce additional empirical balance in particular
$f$-specific directions, in which case undersmoothing of HAL may be paired with
the score-preserving properties of targeted updating to ensure efficiency while
accommodating balance w.r.t.~user-specified functions of $X$.

\section{The Propensity Score's Role in Modern Causal Inference}

The propensity score holds an integral place in causal inference, playing
important roles in terms of both theoretical and empirical balance and
asymptotic efficiency as outlined by developments in semiparametric theory.
We have argued that this empirical balancing property corresponds to solving
particular types of score equations, generally of the form $P_n h(X)(Z
- e_n(X)) \approx 0$, and we have proposed a score test for evaluating the
degree to which a candidate propensity score estimator $e_n(X)$ achieves
balance. By characterizing the empirical balancing property in terms of the
solving of particular score equations, we circumvent the diagnostic strategies
popular today, which focus on checking empirical balance across only a subset of
(necessarily discrete) covariates $X$ or are derived from and reliant upon the
brittle assumptions underlying parametric estimators of $e_0(X)$, e.g., logistic
regression. This form of empirical balance corresponds with approximately
solving a limited set of score equations derived from focusing upon particular
subsets of covariates.

This characterization bridges the gap between satisfaction of the empirical
balancing property and well-studied strategies for asymptotically efficient
estimation based on the solving of efficient score equations (e.g., the
$D_{\text{CAR}}$ component of the EIF). By demonstrating that there need not be
a disconnect between the dual desiderata of empirical balance and asymptotic
efficiency, we reduce enforcing the empirical balancing property (and diagnosing
deviations from it) to the solving of score equations --- achievable within
well-studied frameworks like targeted learning or nonparametric sieve
estimation. We argue that modern propensity score-based estimation strategies in
causal inference should prioritize solving efficient score equations and,
secondarily, aim to solve a variety of other score equations so as to enforce
the empirical balancing property for a large class of functions of the baseline
covariates. Conveniently, satisfaction of the latter criterion may lead to
automatically satisfying the former. This unified focus on efficiency and
empirical balance in terms of score equations of the form $P_n h(X)(Z - e_n(X))
\approx 0$ allows for both criteria to be satisfied without reliance upon
historically popular but unrealistic modeling assumptions whose utility is
severely limited in the complex observational studies common in today's
biomedical, health, and social sciences research.

\acks{%
We thank an anonymous associate editor and two peer referees of
\textit{Biometrics} who inspired our careful consideration of the interplay
between the balancing property and asymptotically efficient estimation in modern
causal inference through their reviews of~\citet{ertefaie2022nonparametric}.
MJvdL was partially supported by a grant from the National Institute of Allergy
and Infectious Diseases (award no.~R01 AI074345).
}

\vskip 0.2in
\bibliography{refs}
\end{document}